\documentclass[article]{aastex62}

\begin{document}

\title{HD 196390: A tight correlation of differential abundances with condensation temperature}

\correspondingauthor{Charles R. Cowley}
\email{cowley@umich.edu}

\author{Charles R. Cowley}
\affiliation{Department of Astronomy, University of Michigan, 
1085 S. University, Ann Arbor, MI 481090-1107}

\author{Donald J. Bord}
\affiliation{Department of Natural Sciences, University of Michigan-Dearborn,
4901 Evergreen Rd., Dearborn, MI 48128}

\author{Kutluay Y\"{u}ce}
\affiliation{Department of Astronomy and Space Sciences,
Faculty of Science \\ University of Ankara, Ankara, TR-06100, Turkey}

\keywords{stars: abundances --- sun: abundances} 

\begin{abstract}
Bedell et al. (2018) give precision differential abundances for 79 
mostly G-dwarf stars. We correct these abundances for Galactic 
chemical evolution in a manner similar to that used by these 
authors but with parameters derived from linear fits to plots of 
[El/H] vs. age in lieu of [El/Fe]. We examine the resulting 
abundances for correlations with the 50\% condensation temperature 
using values from both Lodders (2003) and Wood et al. (2019), and 
compare with the results of Bedell et al. HD 196390 is distinct in 
having the most significant correlation of the 79-star sample. We 
report statistics for a subset of stars with lower significance, 
but of some interest.
\end{abstract}

\section{Introduction\label{sec:intro}} 

One of the most exciting developments in the field of analytical 
stellar spectroscopy is the possible inference of extrasolar  
planetary  properties  from  precision  differential  abundances  
in  solar  twins \citep{mel09}.   Of special 
significance has been the discovery that small abundance 
differences in the sun and some of its twins are correlated with 
the 50\% condensation temperature, $T_{\rm c}$, defined in \citet{lod03}.  
Such correlations have been suggested by numerous workers to be caused 
by processes related to star-planet formation \citep[see][and 
references cited therein]{nisgus18,ram19}. The 
sun itself is believed to be slightly depleted in refractory elements, 
leading to the suggestion that those elements have been sequestered 
in the terrestrial planets.
 
\section{Corrections for Galactic chemical evolution\label{sec:GCE}}

The solar twin HD 196390 (HIP 101905) is one of 79 stars with 
precision differential abundances (PDAs) in the study of \citet[henceforth, 
BD18]{bed18}.  We find its PDA distribution, after 
corrections for Galactic chemical evolution (GCE) effects (see below), 
to exhibit the tightest correlation with condensation temperature of 
any of the stars in this sample.  

BD18 derived GCE parameters $m(Z)$ and $b(Z)$, where $Z$ is the atomic number 
of element $El$, using data for 68 of their 79 stars after removing stars older than
8 Gyr or displaying anomalous abundances for neutron-addition species.  Numerical values for these 
quantities were derived from fits for $[El/Fe] = m(Z) · t + b(Z)$, 
with $t$ the stellar age in Gyr; the results are given in their Tab. 3. \citet{acu20} has derived similar 
GCE parameters for 19 elements based on independent PDA measures in 17 
sun-like stars; for most elements in common with BD18, her values 
for the fitted slopes agree to within the quoted uncertainties with those of BD18.
 
We find empirically that, on average, the function $GCE(t) = m(Z) · t + b(Z)$ 
is an excellent approximation to plots of $[El/H]$ for typical old and 
young stars in the BD18 sample \citep[see][Panel 4, Slides 1 and 2]{cyb20}.  
Thus, for any star in the sample, we  consider
\begin{equation}   
\Delta[El/H] = [El/H] -GCE = [El/H] - \{m(Z)\cdot t + b(Z)\}
\label{eq:GCEcor}
\end{equation}
\noindent to represent the elemental abundance difference between star and sun that {\em is not} 
due to GCE.

We have determined GCE parameters, $m(Z)$ and $b(Z)$, based on $[El/H]$ rather than $[El/Fe]$
as in BD18. The absolute value of the differences between
observed and predicted abundances of the 68-star sample was slightly smaller when
the BD18 coefficients were used instead of those based on $[El/H]$.  This could be because
correlations of $[El/Fe]$ vs. age are generally tighter than those based on $[El/H]$. 

\section{HD 196390 and condensation temperature\label{sec:tccor}}

\begin{figure}[t]                                           
\begin{center}
\includegraphics[scale=0.70,angle=0]{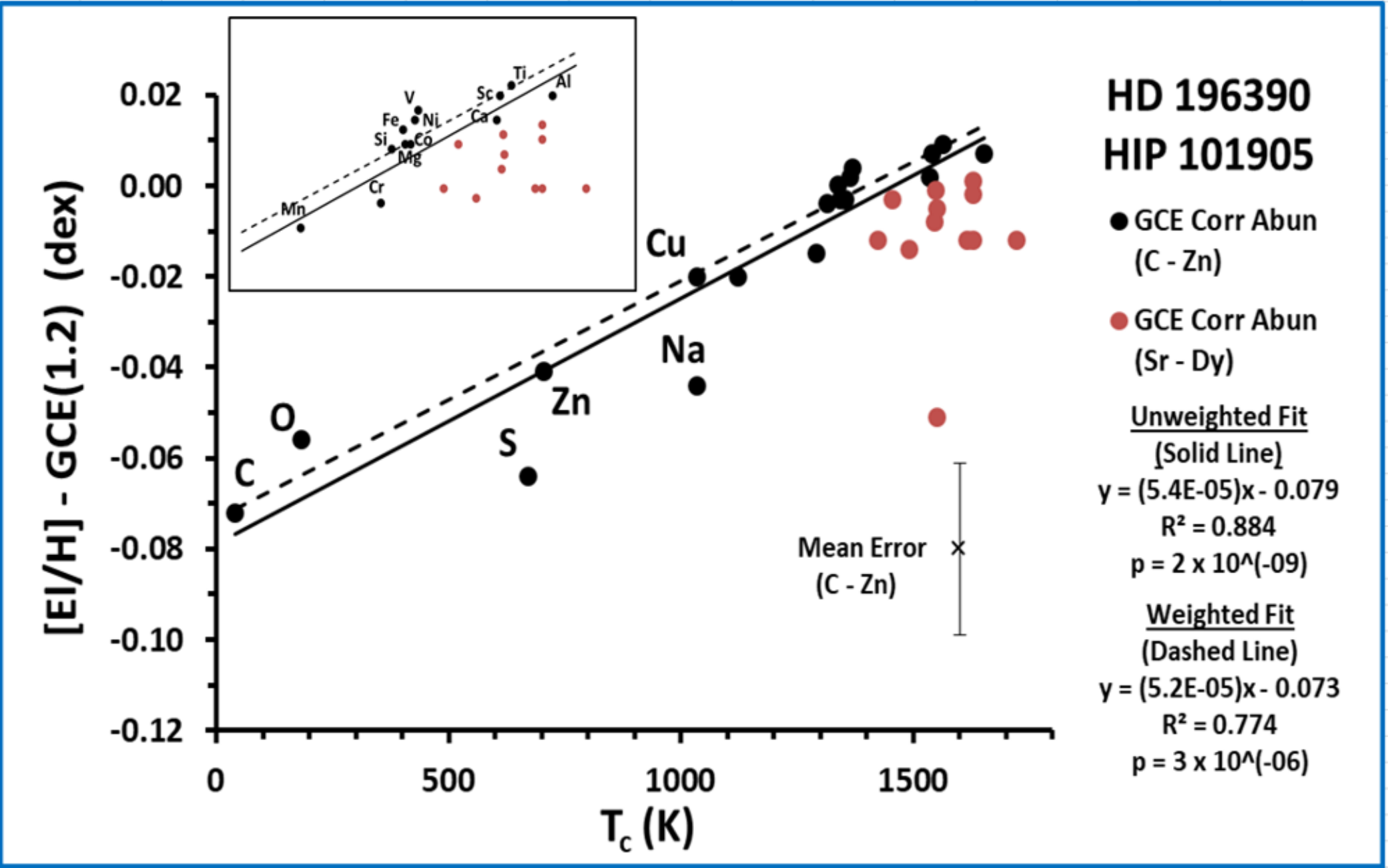}
\caption{Differential abundances for HD 196390 vs. condensation 
temperature, $T_{\rm c}$, after GCE corrections based on the BD18 parameters. 
Values for $T_{\rm c}$ are taken from \citet{wood19}. Black circles designate 
elements C-Zn; gray (red online) circles denote the neutron-addition elements Sr-Dy. 
Linear least-squares fits to the data for C-Zn are shown: The solid line gives the best fit
with all points weighted equally (or essentially unweighted); the dashed line displays the 
best fit with each pointed weighted by its (variance)$^{-1}$ computed using composite 
uncertainties derived by propagating the reported uncertainties in the BD18 PDAs and GCE correction 
parameters, as well as in the star's age in the usual manner assuming the errors to be uncorrelated.
For illustration, the mean error for the fitted elements is shown; errors for the volatile elements 
(C, O, S, Na, Cu, and Zn) are somewhat larger than that shown, while those for the 
remaining elements in the set are smaller.  The statistical significance of either fit 
is very high with {\em p} $\sim$ 10$^{-6}$ or less.
\label{fig:1}}
\end{center}
\end{figure}

The quantity $\Delta[El/H]$ defined in Eq.~\ref{eq:GCEcor} is plotted for 30 elements in HD 196390 in 
Fig.~\ref{fig:1} for an assumed age of 1.2 Gyr \citep{spin18}. 
The quantitative values of the `significance' of the linear fits  
displayed in Fig.~\ref{fig:1}, as measured, for example, by {\em p}, the probability that the 
fit arises by chance, depend on the specific choice of elements, as well as on the method 
of assigning uncertainties and weights for the points included in the fits and the source of
the values of $T_{\rm c}$.    
In Fig.~\ref{fig:1}, we fit {\em only} the elements carbon through zinc. 
Our rationale is that the neutron-addition elements, Sr-Dy, owe at least some of 
their abundance anomalies to processes unrelated to condensation \citep[see][]{mel14}. 
We accept this viewpoint with the caveat that it deserves deeper consideration.  
The adopted uncertainties reflect contributions from the uncertainties in the tabulated 
PDAs for this star, as well as those associated with the parameters $m(Z)$, $b(Z)$, and $t$.  
Several different weighting schemes based on the uncertainties were applied; for all 
plausible choices, we find the significance of the fit for HD 196390 stands well out 
from its congeners in the BD18 sample. Tests of the degree to which the fits are sensitive 
to the sources, either \citet{lod03} or \citet{wood19}, reveal only small differences which do not
alter the overall conclusions presented here (see Appendix).

We have made additional $T_{\rm c}$ fits for all of the other BD18 stars.
These have been also been made using the newly derived GCE parameters based on correlations of 
$[El/H]$ vs. age, as well as the BD18 parameters based on $[El/Fe]$.
We have again compared results from using both the \citet{lod03} and 
\citet{wood19} values of $T_{\rm c}$.  The results for three data sets are presented in 
the Appendix. While they differ in small details, they all point
to the prominence of HD 196390.  In addition, they agree on the stars with the next 
most significant fits, though the ranking by probabilities varies somewhat.  
These stars show formal significances with probabilities in the range $10^{-3}$ 
to $10^{-5}$.  The top (five) stars are: HD 42618 
(HIP 29432), $\psi$ Ser (HD 140538; HIP 77052), HD 45346  (HIP 30502), HD 12264 (HIP 9349),
and HD 158210 (HIP 85402). 
Of these, only HD 42618 (5.2 Gyr) has a known planet.  Interestingly, for this star, the 
slope of the plot  $\Delta[El/H]$ vs. $T_{\rm c}$  is negative.


\citet{mel14} found a very tight correlation of $[El/H]$ with $T_{\rm c}$ for the 
solar twin 18 Sco (HD 146233, HIP 79672), also included in the BD18 sample. Their correlation 
was also developed for the elements carbon through zinc using $T_{\rm c}$ values from
\citet{lod03}, but for PDAs uncorrected for GCE effects. We find that this correlation 
weakens significantly once GCE corrections are applied, regardless of the source of the
correction parameters used in Eq.~\ref{eq:GCEcor}. This was also found by \citet{acu20}, 
who gives a slope after correction of $(0.96\pm 1.61) \cdot 10^{-5}$K$^{-1}$.

While we have made our fits for elements C-Zn, other workers, e.g. BD18, 
sometimes neglect the elements with $T_{\rm c} < 900$K while including the 
{\em n}-addition elements, 
or use a broken-line fit to accommodate the volatiles. The complexities associated with 
star/planet formation likely admit the legitimate application of all these approaches.

\appendix


Tab. 1 presents the statistics for the most significant of the
$\Delta[El/H]$ vs. $T_{\rm c}$ for three sets of input data.  Set A comes from the use
of BD18 GCE parameters according to Eq.~\ref{eq:GCEcor}, and $T_{\rm c}$ of
\citet{lod03}.  Set B used GCE parameters based on fits to $[El/H]$ vs age,
and $T_{\rm c}$ from \citet{lod03}.  Set C used BD18 GCE parameters and the more recent
$T_{\rm c}$ from \citet{wood19}.  
For each star, we give (1) the probability
the fit is due to chance, and (2) the slope of the
$\Delta[El/H]$ vs. $T_{\rm c}$ plots.  The probabilities are all smaller than 
10$^{-4}$.
Note that the order of the stars is nearly
the same in all three sets of results. 

The set of four stars below the horizontal line are again common to all three sets.
They have probabilities between 
10$^{-4}$ and 10$^{-3}$, and are of marginal interest.

\begin{deluxetable}{lcc|lcc|lcc}  [ht]
\tablecaption{Statistics of three T$_{\rm c}$-slope plots}
\tablecolumns{9}
\tablewidth{0pt}
\tablehead{
\multicolumn{3}{c}{Set A} & 
\multicolumn{3}{c}{Set B} &
\multicolumn{3}{c}{Set C}  \\
\colhead{star} & 
\colhead{prob} &
\colhead{slope}& 
\colhead{star} &
\colhead{prob} &
\colhead{slope}&
\colhead{star} &
\colhead{prob} &
\colhead{slope}  
}
\startdata 
HD 196390&	1.96E-09 &  5.38E-05 &HD 196390    &1.19E-07   & 4.90E-05  &HD 196390&1.96E-09& 5.38E-05   \\
HD 42618 &  1.32E-05 & -2.68E-05 &HD 12264    &3.14E-06   &-2.48E-05  &HD 42618 &1.31E-05&-2.68E-05   \\
HD 140538&  2.76E-05 &	5.29E-05 &HD 42618    &6.03E-06   &-2.62E-05  &HD 140538&2.76E-05& 5.29E-05   \\
HD 45346 &	3.76E-05 &  3.74E-05 &HD 45346    &1.73E-05   & 4.04E-05  &HD 45346 &3.76E-05& 3.74E-05   \\
HD 12264 &  5.06E-05 & -1.93E-05 &HD 158210   &2.41E-05   & 3.35E-05  &HD 12264 &5.06E-05&-1.93E-05   \\
HD 158210&	5.94E-05 &  2.94E-05 &HD 140538   &2.68E-05   & 5.29E-05  &HD 158210&5.94E-05& 2.94E-05   \\ \tableline
HD 96423 &	1.05E-04 &  3.51E-05 &HD 96423    &5.97E-05   & 3.75E-05  &HD 96423 &1.05E-04& 3.51E-05   \\
HD 44665 &	1.61E-04 & -1.68E-05 &HD 219057   &1.10E-04   &-3.53E-05  &HD 44665 &1.61E-04&-1.68E-05   \\
HD 134664&	2.50E-04 &  2.51E-05 &HD 68168    &2.91E-04   & 4.23E-05  &HD 134664&2.50E-04& 2.51E-05   \\
HD 68168 &	2.59E-04 &  4.15E-05 &HD 134664   &3.73E-04   & 2.37E-05  &HD 68168 &2.59E-04& 4.15E-05
\enddata
\end{deluxetable}

\acknowledgments
We acknowledge use of the SIMBAD database \citep{wen00} operated at
CDS, Strasbourg, France.
We also used the NASA Exoplanet Archive, which is operated by 
the California Institute of Technology, under contract 
with the National Aeronautics and Space Administration under the 
Exoplanet Exploration Program.

\end{document}